\documentclass[aps,preprint,nofootinbib,letterpaper]{revtex4-2}
\usepackage{amsfonts}
\usepackage{amssymb}
\usepackage{amsmath}
\usepackage[utf8]{inputenc}
\usepackage[colorlinks=true,
            linkcolor=blue,
            urlcolor=blue,
            citecolor=blue]{hyperref}

\begin{document}

\title{Branched Spacetime Geometries from Harmonic $S^2$ Maps: Energy
Matching and Its Limitations}
\author{M. Halilsoy}
\email{mustafa.halilsoy@emu.edu.tr}
\author{S. Habib Mazharimousavi}
\email{habib.mazhari@emu.edu.tr}
\affiliation{Department of Physics, Faculty of Arts and Sciences, Eastern Mediterranean
University, Famagusta, North Cyprus via Mersin 10, T\"{u}rkiye}
\date{\today}

\begin{abstract}
We investigate a topological construction of general-relativistic geometries
based on harmonic self-maps of the two-sphere, $S^{2}\rightarrow S^{2}$.
Such maps are classified by an integer degree $k$. For $k>1$, however, the
pullback angular metric is not globally smooth: it defines a branched
covering with conical excesses at the north and south poles. We calculate
the corresponding distributional curvature and show that the branch points
extend over the $(t,r)$ sector as codimension-two defects with negative
signed tension. For a branched Schwarzschild geometry, we introduce a smooth
Reissner-Nordstr\"{o}m (RN)-like radial deformation and define its finite bulk
Killing energy relative to the $Q_{k}=0$ geometry in the same topological
sector. An explicit phenomenological matching prescription relates this bulk
energy to the scaled excess harmonic-map energy and determines $Q_{k}$.
Identifying the matching scale with the outer horizon then yields a discrete
sequence approaching an extremal limit. We also examine the associated
horizon geometry, conditional entropy, regular curvature scalars, weak-field
limit, and horizonless zero-mass sector. The same positive matching
prescription does not extend universally. For the Simpson-Visser radial
ansatz, the relative bulk Killing energy is negative. In the de Sitter case,
the reduction $\Lambda \rightarrow \Lambda /k$ reproduces the excess map
energy only through a common-volume reference subtraction, not through the
direct energy difference between the physical sectors. The construction
therefore provides a classical framework for branched topological sectors
while also identifying the assumptions and limitations of the associated
energy-matching prescription.
\end{abstract}

\maketitle

\emph{This paper is dedicated to Prof. Metin G\"{u}rses on the occasion of
his completion of sixty fruitful years of contributions to the field of
general relativity.}
\section{Introduction}
Harmonic mapping between two Riemannian manifolds $\mathcal{M}$ and $%
\mathcal{M}^{\prime }$, described by the line elements 
\begin{equation}
ds^{2}=g_{ab}dx^{a}dx^{b},\qquad (a,b=1,2,\ldots ,m),
\end{equation}%
and 
\begin{equation}
ds^{\prime \,2}=g_{AB}dy^{A}dy^{B},\qquad (A,B=1,2,\ldots ,n),
\end{equation}%
respectively, is defined by the energy functional \cite{eells1964}

\begin{equation}
E[y] = \frac{1}{2} \int_{\mathcal{M}} g_{AB}(y) \frac{\partial y^A}{\partial
x^a} \frac{\partial y^B}{\partial x^b} g^{ab}\sqrt{|g|}\,d^m x.
\label{hm-energy}
\end{equation}
in which $|g|=\det g_{ab}$. The extremal condition $\delta E[y^{A}]=0$ yields

\begin{equation}
g^{ab}\left( \nabla _{a}\nabla _{b}y^{A}+\Gamma _{BC}^{A}\frac{\partial y^{B}%
}{\partial x^{a}}\frac{\partial y^{C}}{\partial x^{b}}\right) =0,
\label{hm-equation}
\end{equation}%
where $\nabla _{a}$ is the covariant derivative on $\mathcal{M}$ and $\Gamma
_{BC}^{A}$ are the Christoffel symbols on $\mathcal{M}^{\prime }$. For
appropriate choices of $\mathcal{M}$ and $\mathcal{M}^{\prime }$, these
equations are equivalent to the Einstein field equations \cite%
{nutku1974,misner1978}. The isometries on $\mathcal{M}^{\prime }$ can be
used to generate new solutions from known ones in general relativity \cite%
{halil1981}. The technique of harmonic maps has aided in obtaining important
solutions in the past \cite{nutku1977}. 

Applications of harmonic maps in general relativity fall into several
related but conceptually distinct classes. First, symmetry-reduced vacuum
or Einstein-matter equations can be formulated as geodesic or harmonic-map
systems whose target-space symmetries provide methods for constructing exact
solutions. Early applications included stationary vacuum spacetimes with
whole-cylinder symmetry and other static or stationary gravitational fields
\cite{erisnutku1975,eris1977}. The connection between the Einstein equations
and nonlinear sigma models was subsequently developed for dimensionally
reduced gravitational systems and vacuum spacetimes admitting a spacelike
$U(1)$ isometry \cite{sanchez1982,moncrief1986}. Harmonic-map techniques
were also applied to stationary axisymmetric solutions, higher-dimensional
gravity, and gravitational theories coupled to electromagnetic, dilaton,
and axion fields
\cite{matosplebanski1994,matos1994,matosrios1999,matos2009}.

In the second class, the map is a genuine matter field governed by a
nonlinear sigma-model action and coupled dynamically to gravity. Within
this framework, no-hair and uniqueness theorems have been established for
static and rotating black holes with self-gravitating harmonic mappings
\cite{heusler1992,heusler1995}. Self-gravitating nonlinear sigma models
have also been investigated in the context of critical gravitational
collapse \cite{husa2000}. Harmonic-map ans\"atze have been used to construct
gravitating $SU(N)$ monopoles and Skyrmions
\cite{brihayemonopoles2004,brihaye2004}, while generalized hedgehog
configurations provide further classes of self-gravitating Skyrmion
solutions \cite{canforamaeda2013}. More general aspects of harmonic and
wave maps coupled to Einstein gravity were studied in
Ref.~\cite{schimming2013}. Particularly relevant to the present work are
Einstein--nonlinear-sigma-model solutions with an $S^2$ target, for which
the mass per unit length is proportional to an integer winding number
\cite{bongadotti2022}. More recently, nonlinear sigma models have been used
to construct black holes and black branes with nontrivial horizon
geometries controlled by integer topological invariants
\cite{canforabumpy2026,canfora2026}.

A third line of research employs singular harmonic maps into
nonpositively curved symmetric target spaces. Such methods have played an
important role in establishing regularity and uniqueness results for
stationary black holes \cite{litian1992}, constructing stationary
axisymmetric multi-black-hole solutions of the Einstein-Maxwell equations
\cite{weinstein1996}, and analyzing singular harmonic maps arising in
general-relativistic boundary-value problems \cite{nguyen2011}. Recent
developments have further connected their asymptotic behavior with
near-horizon geometries and mass-angular-momentum inequalities for
multiple-black-hole configurations \cite{han2026}.

The construction considered here differs from these standard applications.
The harmonic map is neither introduced as an independent spacetime matter
field obtained by varying a four-dimensional sigma-model action nor identified
with the target-space potentials arising from a Killing reduction of the
Einstein equations. Instead, a holomorphic self-map $S^{2}\rightarrow S^{2}$
of degree $k$ is pulled back directly into the angular sector of a seed metric.
For $k>1$, the resulting geometry is a branched covering with distributional
polar defects. The excess Dirichlet energy is then related to a smooth radial
source through an additional phenomenological matching condition. Thus, the
integer $k$ has a geometric and topological origin, whereas the corresponding
mass shift is model-dependent rather than a consequence of a covariant
Einstein--sigma-model action.

The particular importance of maps
from $S^{2}$ to $S^{2}$ follows from their topological classification. A
smooth map 
\begin{equation}
\mathbf{n}:S^{2}\longrightarrow S^{2},\qquad \mathbf{n}\cdot \mathbf{n}=1,
\end{equation}%
is characterized by its degree 
\begin{equation}
k=\frac{1}{4\pi }\int_{S^{2}}\mathbf{n}\cdot \left( \partial _{\theta }%
\mathbf{n}\times \partial _{\varphi }\mathbf{n}\right) d\theta \,d\varphi
\in \mathbb{Z}.  \label{map-degree}
\end{equation}%
This integer counts, with orientation, the number of times the domain sphere
wraps around the target sphere. The relevant topological invariant is the
second homotopy group $\pi _{2}(S^{2})$, defined as the set of homotopy
classes of continuous maps from the two-sphere into itself. For $S^{2}$,
this group is isomorphic to the integers: 
\begin{equation}
\pi _{2}(S^{2})\cong \mathbb{Z}.
\end{equation}%
Consequently, two maps with different degrees cannot be continuously
deformed into one another without leaving their respective homotopy classes
or passing through a singular configuration. The configuration space
therefore separates into distinct topological sectors labeled by $k$. For
the nonlinear sigma-model energy, the degree provides the topological lower
bound 
\begin{equation}
E[\mathbf{n}]\geq 4\pi |k|.  \label{topological-bound}
\end{equation}%
Holomorphic or antiholomorphic maps between Riemann spheres saturate this
bound and are consequently harmonic. In terms of the stereographic
coordinate 
\begin{equation}
z=\tan \left( \frac{\theta }{2}\right) e^{i\varphi },  \label{R1}
\end{equation}%
these solutions are represented by rational functions $R(z)$. In particular,
the map 
\begin{equation}
R(z)=z^{k}  \label{R2}
\end{equation}%
has degree $k$ and gives the angular factor $f_{k}(\theta )$ employed below.
Thus, the discreteness of $k$ is not imposed algebraically but follows from
the homotopy classification of maps between two-spheres.

Topological maps of this type play an important role in several areas of
physics. In the two-dimensional $O(3)$ nonlinear sigma model, the degree of
the map is the topological charge of the Belavin-Polyakov soliton,
describing a localized magnetic configuration protected against continuous
decay \cite{belavin1975}. Rational maps between Riemann spheres have also
been used to construct and classify magnetic monopoles and multisoliton
configurations in the Skyrme model, where their degree is related to the
monopole or baryon number \cite{houghton1998}. Harmonic-map ans\"{a}tze have
further been applied to gravitating Skyrmions in Einstein-Skyrme systems,
demonstrating that the same topological methods remain useful when nonlinear
sigma-model fields are coupled to gravity \cite{brihaye2004}.

More recent developments have extended these ideas in several directions.
The existence and stability of rotationally symmetric $p$-harmonic self-maps
of spheres have been investigated in Ref.~\cite{branding2023}.
Generalizations of the rational-map construction to compact Hermitian
symmetric spaces have been obtained from self-duality equations, opening
further applications to nonlinear sigma models, Skyrme theories, and
Yang-Mills-Higgs monopoles \cite{ferreira2024}. Rational maps have also
acquired direct experimental relevance in structured optics, where they are
used to generate optical Skyrmions, antiskyrmions, bimerons, and
multiskyrmion configurations \cite{cisowski2023}. In gravitation, nonlinear
sigma models continue to provide a productive framework for constructing
nontrivial black-hole, black-string, and cosmological geometries \cite%
{canfora2026}. These applications illustrate a common mechanism: a
topological integer becomes physically meaningful when it is connected to an
energy, conserved charge, or geometrical quantity. In the present
construction, the degree $k$ first enters through the harmonic map of the
angular two-sphere. The corresponding map energy is then matched to the
energy-momentum source supporting the modified radial geometry. In this way,
the term $Q_{k}^{2}/r^{2}$ introduced below is not an independent
RN charge; rather, its coefficient is fixed by requiring
the integrated source energy to reproduce the excess harmonic-map energy
relative to the vacuum sector $k=1$.

In this work, rather than considering arbitrary manifolds, we restrict
ourselves to harmonic maps between two-spheres. Our technique of geometric
discretization is built upon a well-known classical spacetime such that, for
the discrete parameter $k=1$, the continuum is recovered. To expose the
power of harmonic maps, symmetry, and the natural discreteness they induce
in classical geometry, we study four classes of metrics:

\begin{enumerate}
\item The branched Schwarzschild geometry, in which the cutoff and matching
scale $\ell_h$ is fixed self-consistently by identifying it with the outer
horizon radius. The smooth $Q_k^2/r^2$ contribution is then determined by
matching its relative bulk Killing energy to the scaled excess harmonic-map
energy.

\item The horizonless zero-mass geometry, which retains the smooth $Q_k$%
-dependent bulk source and the two branch defects but possesses a naked
curvature singularity at $r=0$. An independent inner cutoff $L_0$ may be
imposed to excise the singular region; identifying it with the Planck length
is an additional microscopic assumption.

\item The Simpson-Visser geometry, for which the harmonic map produces the
same branched angular sector. A direct calculation of the complete Einstein
tensor shows, however, that the positive bulk-energy matching used in the
Schwarzschild case does not extend to the RN-like radial
deformation of the wormhole. This example identifies a limitation of the
matching prescription.

\item The de Sitter geometry, in which the regular cosmological constant and
scalar curvature scale as $\Lambda_k=\Lambda/k$ and $R_k=4\Lambda_k$. In a
common coordinate region, the direct regular vacuum energies of the $k$ and $%
k=1$ sectors are equal because the decrease of the vacuum density is
compensated by the increased branched angular volume. The excess map energy
is reproduced only by a reference subtraction in which both densities are
evaluated on the same $k$-branched volume.
\end{enumerate}

This work extends Ref.~\cite{halilsoy2023} in several essential
respects. First, we analyze the global structure of the harmonic-map
pullback and identify its branch points, conical excesses, and
distributional energy-momentum tensor. Second, we distinguish the divergent
defect energy from the finite relative Killing energy of the smooth bulk
source and formulate the latter matching with an explicit normalization
parameter. Third, we determine the horizon scale self-consistently and
examine the thermodynamic and weak-field consequences under clearly stated
assumptions. Finally, the Simpson-Visser and de Sitter examples establish
that the angular topological construction is more general than the proposed
bulk-energy matching, which must be tested independently for each radial
geometry.

\section{The Schwarzschild black hole}

All static, spherically symmetric $(3+1)$-dimensional black-hole geometries
contain an angular $S^{2}$ sector. A harmonic self-map of this sector
introduces an integer topological degree $k$, representing the number of
wrappings of the domain sphere around the target sphere. We use this degree
to label the corresponding branched geometries. We adopt units in which $%
c=\hbar =1$ and define $\kappa \equiv 8\pi G,$ restoring SI units where
required. Applying the harmonic-map pullback to the angular sector gives the
following line element, which is locally Schwarzschild away from the branch
points

\begin{equation}
ds^{2} = -\left(1-\frac{2m}{r}\right)dt^{2} + \frac{dr^{2}}{1-\frac{2m}{r}}
+ r^{2}f_{k}^{2}(\theta)d\Omega^{2},  \label{mapped-schwarzschild}
\end{equation}
where

\begin{equation}
f_{k}(\theta) = \frac{ 2k(\sin\theta)^{k-1} }{ (1-\cos\theta)^{k} +
(1+\cos\theta)^{k} },  \label{fk-definition}
\end{equation}
and

\begin{equation}
d\Omega ^{2}=d\theta ^{2}+\sin ^{2}\theta \,d\varphi ^{2}.
\label{unit-sphere}
\end{equation}%
The resulting line element is locally isometric to the Schwarzschild
geometry away from the poles, while its global branched structure will be
examined below. Here $k\in\mathbb{Z}$ measures the degree of the harmonic
map on $S^{2}$ \cite{eells1964}. The
energy of the map is $E_{k}=4\pi k$, and in what follows, we restrict
ourselves to $k\in \mathbb{N}$. It is important to emphasize that this
restriction is not merely an algebraic assumption but follows from the
global topology of the map. In stereographic coordinates, the relevant
harmonic map is represented by (\ref{R1}) and (\ref{R2}) such that under the
periodic identification $\varphi \sim \varphi +2\pi $, this map transforms
according to 
\begin{equation}
R(z)\longrightarrow e^{2\pi ik}R(z).  \label{R3}
\end{equation}

\subsection{Global structure of the mapped two-sphere}

Equation~(\ref{R3}) shows that global single-valuedness requires $e^{2\pi
ik}=1$ and hence $k\in \mathbb{Z}$. Restricting to orientation-preserving
maps gives $k\in \mathbb{N}$, so that $k$ is the well-defined degree of the
map. For noninteger $k$, the map is multivalued and requires a branch cut.
It therefore does not define a global map $S^{2}\rightarrow S^{2}$. This
global single-valuedness must, however, be distinguished from the regularity
of the pullback metric.

Writing the target-sphere coordinates as $(\Theta ,\Phi )$, the map $%
R(z)=z^{k}$ is equivalently expressed as 
\begin{equation}
\tan \left( \frac{\Theta }{2}\right) =\tan ^{k}\left( \frac{\theta }{2}%
\right) ,\qquad \Phi =k\varphi .
\end{equation}%
Away from the branch points, its pullback satisfies 
\begin{equation}
d\Theta ^{2}+\sin ^{2}\Theta \,d\Phi ^{2}=f_{k}^{2}(\theta )\left( d\theta
^{2}+\sin ^{2}\theta \,d\varphi ^{2}\right) .
\end{equation}%
Hence, the mapped angular geometry is locally isometric to the unit
two-sphere wherever the differential of the map is nonvanishing. For integer 
$k>1$, however, the map is a branched covering rather than a global
diffeomorphism. As $\theta \rightarrow 0$, 
\begin{equation}
f_{k}(\theta )\simeq \frac{k}{2^{k-1}}\theta ^{k-1}.
\end{equation}%
Upon defining 
\begin{equation}
\rho =\frac{\theta ^{k}}{2^{k-1}},
\end{equation}%
the angular line element becomes 
\begin{equation}
f_{k}^{2}(\theta )d\Omega ^{2}\simeq d\rho ^{2}+k^{2}\rho ^{2}d\varphi ^{2}.
\end{equation}%
Since $\varphi $ has period $2\pi $, a small circle around the north pole
has circumference $2\pi k\rho $. The corresponding signed conical deficit is
therefore 
\begin{equation}
\delta _{N}=2\pi (1-k)<0,
\end{equation}%
or, equivalently, the pole has a conical excess of magnitude $2\pi (k-1)$.
The same analysis near $\theta =\pi $ gives 
\begin{equation}
\delta _{S}=2\pi (1-k)<0.
\end{equation}

The distributional contributions at the branch points are required by the
Gauss-Bonnet theorem. Away from the poles, the mapped angular metric has
Gaussian curvature $K=1$, while its area is 
\begin{equation}
\int_{S^{2}}f_{k}^{2}(\theta )\,d\Omega =4\pi k.
\end{equation}%
Consequently, 
\begin{equation}
\int_{S^{2}\setminus \{N,S\}}K\,dA+\delta _{N}+\delta _{S}=4\pi k+2\pi
(1-k)+2\pi (1-k)=4\pi =2\pi \chi (S^{2}),
\end{equation}%
as required for $\chi (S^{2})=2$. Thus, the conical contributions are
necessary for the global curvature of the branched two-sphere to remain
consistent with its Euler characteristic. The mapped geometry must therefore
be interpreted globally as a branched-cover geometry. It is smooth and
locally isometric to the original angular geometry away from the two branch
points, whereas the branch points carry distributional curvature. For $k=1$,
both conical contributions vanish and the ordinary smooth two-sphere is
recovered. For $k>1$, the north and south branch points extend over the $%
(t,r)$ sector of the four-dimensional spacetime and represent
codimension-two radial defects. Their distributional contribution to the
Einstein equations must be distinguished from the smooth energy-momentum
tensor generated by the term $Q_{k}^{2}/r^{2}$ considered below.

\subsection{Distributional source of the branch defects}

\label{sec:branch-defects}

The conical excesses at the north and south poles produce distributional
contributions to the four-dimensional Einstein tensor. To identify them, let 
\begin{equation}
d\Sigma _{k}^{2}=q_{ij}dx^{i}dx^{j}=f_{k}^{2}(\theta )\left( d\theta
^{2}+\sin ^{2}\theta \,d\varphi ^{2}\right)
\end{equation}%
denote the mapped angular metric, and introduce angular delta distributions
normalized according to 
\begin{equation}
\int_{S^{2}}\delta _{N}^{(2)}\,dA_{q}=\int_{S^{2}}\delta
_{S}^{(2)}\,dA_{q}=1,
\end{equation}%
where 
\begin{equation}
dA_{q}=f_{k}^{2}(\theta )\sin \theta \,d\theta \,d\varphi .
\end{equation}%
The scalar curvature of the mapped two-sphere, including its distributional
part, is then 
\begin{equation}
{}^{(2)}R=2+4\pi (1-k)\left[ \delta _{N}^{(2)}+\delta _{S}^{(2)}\right] .
\label{angular-distributional-curvature}
\end{equation}%
Indeed, the regular part integrates to $8\pi k$, while the two
distributional terms contribute $8\pi (1-k)$, giving 
\begin{equation}
\int_{S^{2}}{}^{(2)}R\,dA_{q}=8\pi ,
\end{equation}%
in agreement with the Gauss-Bonnet theorem. Consider now a four-dimensional
warped-product line element of the form 
\begin{equation}
ds^{2}=h_{AB}(x)\,dx^{A}dx^{B}+r^{2}(x)q_{ij}(y)\,dy^{i}dy^{j},\qquad A,B\in
\{t,r\}.
\end{equation}%
The distributional part of the four-dimensional scalar curvature is obtained
directly from Eq.~\eqref{angular-distributional-curvature}: 
\begin{equation}
R_{\mathrm{def}}=\frac{4\pi (1-k)}{r^{2}}\left[ \delta _{N}^{(2)}+\delta
_{S}^{(2)}\right] .
\end{equation}%
The $(t,r)$ components of the Ricci tensor contain no corresponding
distributional term. It follows that the nonvanishing mixed components of
the distributional Einstein tensor are 
\begin{equation}
\left( G^{A}{}_{B}\right) _{\mathrm{def}}=-\frac{2\pi (1-k)}{r^{2}}\left[
\delta _{N}^{(2)}+\delta _{S}^{(2)}\right] \delta ^{A}{}_{B},\qquad A,B\in
\{t,r\},  \label{defect-einstein-tensor}
\end{equation}%
whereas 
\begin{equation}
\left( G^{i}{}_{j}\right) _{\mathrm{def}}=0.
\end{equation}%
Adopting the convention 
\begin{equation}
G^{\mu }{}_{\nu }=\kappa T^{\mu }{}_{\nu },\qquad \kappa =8\pi G,
\end{equation}%
the effective energy-momentum tensor of the two branch defects is therefore 
\begin{equation}
\left( T^{A}{}_{B}\right) _{\mathrm{def}}=-\frac{2\pi (1-k)}{\kappa r^{2}}%
\left[ \delta _{N}^{(2)}+\delta _{S}^{(2)}\right] \delta ^{A}{}_{B}.
\label{defect-energy-momentum}
\end{equation}%
Equivalently, each radial defect has the signed energy per unit radial
length 
\begin{equation}
\mu _{k}=\frac{\delta _{N}}{\kappa }=\frac{1-k}{4G}.
\end{equation}%
Thus, $\mu _{k}<0$ for $k>1$, as expected for a conical excess rather than a
conical deficit. In mixed-component notation, the corresponding
distributional energy density and radial pressure satisfy 
\begin{equation}
\rho _{\mathrm{def}}=\frac{2\pi (1-k)}{\kappa r^{2}}\left[ \delta
_{N}^{(2)}+\delta _{S}^{(2)}\right] <0,\qquad p_{r,\mathrm{def}}=-\rho _{%
\mathrm{def}},
\end{equation}%
while the angular distributional pressures vanish. The branch defects
therefore saturate the radial null-energy condition but violate it for null
directions having an angular component. Because the defects extend over the
entire radial sector, their absolute Killing energy is not finite in an
asymptotically unbounded spacetime. For a static metric satisfying $%
-g_{tt}g_{rr}=1$, their energy between $r=r_{1}$ and $r=r_{2}$ is 
\begin{equation}
E_{\mathrm{def}}(r_{1},r_{2})=2\mu _{k}(r_{2}-r_{1})=\frac{1-k}{2G}%
(r_{2}-r_{1}),
\end{equation}%
where the factor of two accounts for the north and south defects.
Consequently, 
\begin{equation}
\lim_{r_{2}\rightarrow \infty }E_{\mathrm{def}}(r_{1},r_{2})=-\infty \qquad
(k>1).
\end{equation}%
The branched geometries are therefore not globally asymptotically flat in
the ordinary sense, even when their local metric functions approach those of
an asymptotically flat spacetime.

In what follows, the energy associated with the smooth $Q_k^2/r^2$ source
must consequently be understood as a bulk energy measured relative to the $%
Q_k=0$ branched geometry with the same value of $k$. Under this subtraction,
the identical distributional defect contributions cancel. The matching
condition introduced below applies only to this finite bulk-energy
difference and not to the absolute energy of the complete branched spacetime.

\subsection{The branched Schwarzschild geometry and bulk-energy matching}

To elevate the continuous geometry to a discrete geometry, we employ the
integer $k>1$ as a discrete number and modify the Schwarzschild metric as

\begin{equation}
ds^{2} = -g(r,k)dt^{2} + \frac{dr^{2}}{g(r,k)} +
r^{2}f_{k}^{2}(\theta)d\Omega^{2},  \label{discrete-schwarzschild}
\end{equation}
where

\begin{equation}
g(r,k)=1-\frac{2m}{r}+\frac{Q_{k}^{2}}{r^{2}}.  \label{metric-function}
\end{equation}%
Here $Q_{k}$ is initially an undetermined parameter characterizing the
smooth bulk modification of the radial metric function. It should not be
confused with an electromagnetic charge. Its dependence on the topological
degree $k$ will be fixed below by an explicit energy-matching prescription.
Away from the two branch defects, the $Q_{k}$-dependent part of the mixed
Einstein tensor is 
\begin{equation}
\left( G^{\mu }{}_{\nu }\right) _{\mathrm{bulk}}
=\frac{Q_{k}^{2}}{r^{4}}\operatorname{diag}(-1,-1,1,1).
\label{bulk-einstein-tensor}
\end{equation}%
Using 
\begin{equation}
G^{\mu }{}_{\nu }=\kappa T^{\mu }{}_{\nu },\qquad \kappa =8\pi G,
\end{equation}%
the corresponding smooth bulk energy-momentum tensor is 
\begin{equation}
\left( T^{\mu }{}_{\nu }\right) _{\mathrm{bulk}}
=\frac{Q_{k}^{2}}{\kappa r^{4}}\operatorname{diag}(-1,-1,1,1).
\label{bulk-energy-momentum}
\end{equation}%
Thus, 
\begin{equation}
\rho _{\mathrm{bulk}}=\frac{Q_{k}^{2}}{\kappa r^{4}},\qquad p_{r,\mathrm{bulk%
}}=-\rho _{\mathrm{bulk}},\qquad p_{\theta ,\mathrm{bulk}}=p_{\varphi ,%
\mathrm{bulk}}=\rho _{\mathrm{bulk}}.
\end{equation}%
This tensor is distinct from the distributional source of the branch defects
obtained in the preceding subsection. To define the bulk energy
unambiguously, let 
\begin{equation}
\xi ^{\mu }=(\partial _{t})^{\mu }
\end{equation}%
be the static Killing vector and let $n^{\mu }$ be the future-directed unit
normal to a hypersurface $\Sigma _{t}$ of constant $t$. The conserved
Killing energy of the smooth bulk source is 
\begin{equation}
E_{\mathrm{bulk}}(k)=\int_{\Sigma _{t}}\left( T_{\mu \nu }\right) _{\mathrm{%
bulk}}n^{\mu }\xi ^{\nu }\sqrt{h}\,d^{3}x.  \label{bulk-killing-energy}
\end{equation}%
For the metric under consideration, 
\begin{equation}
n^{\mu }=\frac{1}{\sqrt{g(r,k)}}\,\delta _{t}^{\mu },\qquad \sqrt{h}=\frac{%
r^{2}f_{k}^{2}(\theta )\sin \theta }{\sqrt{g(r,k)}}.
\end{equation}%
The lapse factor in $T_{\mu \nu }n^{\mu }\xi ^{\nu }$ cancels the factor $1/%
\sqrt{g(r,k)}$ in the spatial volume element. Consequently, using 
\begin{equation}
\int_{S^{2}}f_{k}^{2}(\theta )\,d\Omega =4\pi k,
\end{equation}%
the bulk Killing energy outside a lower radius $\ell _{h}$ becomes 
\begin{equation}
E_{\mathrm{bulk}}(k)=\frac{Q_{k}^{2}}{\kappa }\int_{\ell _{h}}^{\infty }%
\frac{dr}{r^{2}}\int_{S^{2}}f_{k}^{2}(\theta )\,d\Omega =\frac{4\pi
kQ_{k}^{2}}{\kappa \ell _{h}}.  \label{bulk-energy-result}
\end{equation}%
This is a finite relative energy. It is the difference between the geometry
with $Q_{k}\neq 0$ and the branched reference geometry with $Q_{k}=0$ at the
same value of $k$. The distributional energies of the identical branch
defects cancel in this comparison. The harmonic-map energy 
\begin{equation}
E_{k}^{(\mathrm{map})}=4\pi k
\end{equation}%
is dimensionless. To associate it with a physical energy, a length scale and
a normalization must be supplied. We therefore introduce a dimensionless
coupling $\gamma >0$ and define the physical excess map energy relative to
the $k=1$ sector by 
\begin{equation}
\Delta E_{k}^{(\mathrm{map})}=\frac{\gamma \ell _{h}}{\kappa }\left[ E_{k}^{(%
\mathrm{map})}-E_{1}^{(\mathrm{map})}\right] =\frac{4\pi \gamma (k-1)\ell
_{h}}{\kappa }.  \label{scaled-map-energy}
\end{equation}%
The factor $\ell _{h}/\kappa $ has the dimension of energy in units $c=\hbar
=1$. We now impose the phenomenological matching condition 
\begin{equation}
E_{\mathrm{bulk}}(k)=\Delta E_{k}^{(\mathrm{map})}.
\label{energy-matching-condition}
\end{equation}%
Equations~\eqref{bulk-energy-result} and \eqref{scaled-map-energy} then give 
\begin{equation}
Q_{k}^{2}=\gamma \left( 1-\frac{1}{k}\right) \ell _{h}^{2}.
\label{Qk-general}
\end{equation}%
The choice $\gamma =1$ defines the minimal normalization adopted in the
remainder of this work, for which 
\begin{equation}
Q_{k}^{2}=\left( 1-\frac{1}{k}\right) \ell _{h}^{2}.  \label{Qk-minimal}
\end{equation}%
The relation above is therefore not a consequence of the Einstein equations
alone. It is an additional matching prescription connecting the finite bulk
source energy to the excess harmonic-map energy. A derivation of the
normalization $\gamma $ would require an underlying four-dimensional action
coupling the harmonic-map sector to gravity.

\subsubsection{Self-consistent determination of the cutoff scale}

The lower limit $\ell _{h}$ in the bulk Killing-energy integral should not
be interpreted as a universal fundamental length. It is a sector-dependent
cutoff and matching scale specifying the exterior region over which the
smooth bulk-source energy is evaluated. Because the present construction
concerns the exterior of a black hole, the geometrically preferred lower
boundary is the outer Killing horizon. We therefore impose the
self-consistency condition 
\begin{equation}
\ell _{h}=r_{+}.  \label{cutoff-horizon-condition}
\end{equation}%
This is an additional physical choice such that neither the topology of the
harmonic map nor the local Einstein equations alone requires it. To retain
the dependence on the matching normalization, define 
\begin{equation}
\beta _{k}=\gamma \left( 1-\frac{1}{k}\right) ,
\end{equation}%
so that Eq.~\eqref{Qk-general} becomes 
\begin{equation}
Q_{k}^{2}=\beta _{k}\ell _{h}^{2}.  \label{Qk-beta}
\end{equation}%
The horizon equation 
\begin{equation}
g(r_{+},k)=1-\frac{2m}{r_{+}}+\frac{Q_{k}^{2}}{r_{+}^{2}}=0
\end{equation}%
together with Eqs.~\eqref{cutoff-horizon-condition} and \eqref{Qk-beta}
gives 
\begin{equation}
\ell _{h}=\frac{2m}{1+\beta _{k}},\qquad Q_{k}=\frac{2m\sqrt{\beta _{k}}}{%
1+\beta _{k}}.  \label{general-cutoff-charge}
\end{equation}%
For $\ell _{h}$ to coincide with the outer, rather than the inner, horizon,
one must have 
\begin{equation}
0\leq \beta _{k}\leq 1.  \label{beta-condition}
\end{equation}%
If the same value of $\gamma $ is to be used for every $k\in \mathbb{N}$, a
sufficient condition is $0<\gamma \leq 1$. For the minimal normalization $%
\gamma =1$ adopted in this work, 
\begin{equation}
\beta _{k}=1-\frac{1}{k},
\end{equation}%
and Eq.~\eqref{general-cutoff-charge} reduces to 
\begin{equation}
\ell _{h}=r_{+}=\frac{2k}{2k-1}\,m,\qquad Q_{k}=\frac{2\sqrt{k(k-1)}}{2k-1}%
\,m.  \label{minimal-cutoff-charge}
\end{equation}%
Consequently, 
\begin{equation}
\alpha _{k}\equiv \frac{Q_{k}}{m}=\frac{2\sqrt{k(k-1)}}{2k-1}.
\end{equation}%
The $k=1$ sector gives 
\begin{equation}
Q_{1}=0,\qquad \ell _{h}=r_{+}=2m,
\end{equation}%
and therefore recovers the Schwarzschild horizon. As $k\rightarrow \infty $, 
\begin{equation}
Q_{k}\rightarrow m,\qquad \ell _{h}\rightarrow m,
\end{equation}%
so the sequence approaches the extremal RN-like limit
without exceeding it. The identification $\ell _{h}=r_{+}$ also makes clear
why the divergence of the formal bulk-energy integral at $r=0$ is irrelevant
to the exterior construction. The integration region begins at the horizon
and never includes the central singularity. The cutoff does not remove or
resolve that singularity; it restricts the energy calculation to the
black-hole exterior.

\subsubsection{Horizon geometry and thermodynamic quantities}

The outer horizon is located at 
\begin{equation}
r_{+}=m+\sqrt{m^{2}-Q_{k}^{2}}=m(1+\chi _{k}),  \label{outer-horizon}
\end{equation}%
where 
\begin{equation}
\chi _{k}=\sqrt{1-\frac{Q_{k}^{2}}{m^{2}}}.  \label{chi-definition}
\end{equation}%
Although the horizon cross-section contains conical branch points at its
north and south poles, its total area is finite. Using 
\begin{equation}
\int_{S^{2}}f_{k}^{2}(\theta )\,d\Omega =4\pi k,
\end{equation}%
we obtain 
\begin{equation}
A_{k}=r_{+}^{2}\int_{S^{2}}f_{k}^{2}(\theta )\,d\Omega =4\pi kr_{+}^{2}=4\pi
m^{2}k(1+\chi _{k})^{2}.  \label{branched-horizon-area}
\end{equation}%
Thus, the factor of $k$ in the area is a direct consequence of the $k$-fold
branched angular geometry. The surface gravity associated with the Killing
vector $\xi ^{\mu }=(\partial _{t})^{\mu }$ is determined entirely by the
radial metric function 
\begin{equation}
\varkappa _{H,k}=\frac{1}{2}\left. \frac{\partial g(r,k)}{\partial r}%
\right\vert _{r=r_{+}}=\frac{\sqrt{m^{2}-Q_{k}^{2}}}{r_{+}^{2}}.
\label{surface-gravity}
\end{equation}%
The corresponding Hawking temperature is therefore 
\begin{equation}
T_{H,k}=\frac{\varkappa _{H,k}}{2\pi }=\frac{\chi _{k}}{2\pi m(1+\chi
_{k})^{2}}.  \label{hawking-temperature}
\end{equation}%
Because the branch defects modify the global angular geometry but not the
local $(t,r)$ geometry, they alter the horizon area but do not change the
local surface-gravity formula. For the minimal matching normalization $%
\gamma =1$, Eq.~\eqref{minimal-cutoff-charge} gives 
\begin{equation}
\frac{Q_{k}}{m}=\frac{2\sqrt{k(k-1)}}{2k-1},
\end{equation}%
and hence 
\begin{equation}
\chi _{k}=\frac{1}{2k-1}.
\end{equation}%
The horizon radius, area, and temperature can then be written explicitly as 
\begin{equation}
r_{+}=\frac{2k}{2k-1}\,m,  \label{discrete-horizon-radius}
\end{equation}%
\begin{equation}
A_{k}=\frac{16\pi m^{2}k^{3}}{(2k-1)^{2}},  \label{discrete-horizon-area}
\end{equation}%
and 
\begin{equation}
T_{H,k}=\frac{2k-1}{8\pi mk^{2}}.  \label{discrete-hawking-temperature}
\end{equation}%
For $k=1$, these expressions reduce to the Schwarzschild results 
\begin{equation}
r_{+}=2m,\qquad A_{1}=16\pi m^{2},\qquad T_{H,1}=\frac{1}{8\pi m}.
\end{equation}%
In the limit $k\rightarrow \infty $, 
\begin{equation}
r_{+}\rightarrow m,\qquad \frac{Q_{k}}{m}\rightarrow 1,\qquad
T_{H,k}\rightarrow 0,
\end{equation}%
showing that the sequence approaches an extremal RN-like
geometry. For the Einstein-Hilbert action, and provided that the branch
defects are treated as fixed external sources whose action has no explicit
dependence on the Riemann tensor, the Wald entropy retains the
Bekenstein-Hawking form 
\begin{equation}
S_{k}=\frac{A_{k}}{4G}=\frac{4\pi m^{2}k^{3}}{G(2k-1)^{2}}.
\label{conditional-entropy}
\end{equation}%
This entropy formula should be understood conditionally. A microscopic
action for the branch defects could introduce additional entropy
contributions, particularly if their tension is allowed to vary. These
expressions generalize the local thermodynamic quantities considered in Ref.~%
\cite{halilsoy2023}, now with the global branch structure made explicit. We
emphasize that the usual asymptotically flat RN first
law cannot be imported without modification. The present spacetime is
asymptotically branched and contains radial defects with the signed tension 
\begin{equation}
\mu _{k}=\frac{1-k}{4G}.
\end{equation}%
A complete first law would therefore require properly defined global
Hamiltonian charges and, if variations between different values of $k$ are
permitted, an additional work term associated with the variation of the
defect tension or conical excess. Since $k$ labels disconnected topological
sectors, we restrict the thermodynamic expressions above to a fixed-$k$
sector and do not formulate a first law involving variations of $k$.

\subsubsection{Topological \texorpdfstring{$k$}{k}-dependence of the regular
curvature}

We next examine the $k$-dependence of the regular curvature away from the
north and south branch defects. In a Newman-Penrose tetrad adapted to the
static geometry, the nonvanishing regular Weyl and Ricci scalars are \cite%
{newman1962} 
\begin{equation}
\Psi _{2}^{(\mathrm{reg})}(r,k)=-\frac{m}{r^{3}}+\frac{Q_{k}^{2}}{r^{4}},
\label{regular-Psi2}
\end{equation}%
and 
\begin{equation}
\Phi _{11}^{(\mathrm{reg})}(r,k)=\frac{Q_{k}^{2}}{2r^{4}}.
\label{regular-Phi11}
\end{equation}%
These expressions coincide locally with those of a RN
geometry, although $Q_{k}$ is not an electromagnetic charge in the present
construction. For $k=1$, one has $Q_{1}=0$, and therefore 
\begin{equation}
\Psi _{2}^{(\mathrm{reg})}(r,1)=-\frac{m}{r^{3}},\qquad \Phi _{11}^{(\mathrm{%
reg})}(r,1)=0,
\end{equation}%
as expected for the Schwarzschild geometry. For $k>1$, the $Q_{k}$-dependent
term contributes to both the regular Weyl and Ricci curvatures and falls off
as $r^{-4}$. It is therefore subleading to the Schwarzschild Weyl term at
large radius. The relative magnitude of the two terms in Eq.~%
\eqref{regular-Psi2} is 
\begin{equation}
\mathcal{R}_{k}(r)=\frac{Q_{k}^{2}/r^{4}}{m/r^{3}}=\frac{Q_{k}^{2}}{mr}.
\label{curvature-ratio}
\end{equation}%
Since $r\geq r_{+}$ in the exterior region, this ratio satisfies 
\begin{equation}
\mathcal{R}_{k}(r)\leq \mathcal{R}_{k}(r_{+}).
\end{equation}%
For the minimal normalization $\gamma =1$, Eqs.~\eqref{minimal-cutoff-charge}
and \eqref{discrete-horizon-radius} give 
\begin{equation}
\mathcal{R}_{k}(r_{+})=\frac{2(k-1)}{2k-1}<1\qquad (k<\infty ).
\label{horizon-curvature-ratio}
\end{equation}%
Thus, the smooth $Q_{k}$-dependent contribution never dominates the
Schwarzschild Weyl term anywhere in the exterior of a finite-$k$ black hole.
It is largest at the outer horizon and approaches the same magnitude as the
Schwarzschild term only in the limiting extremal sector, 
\begin{equation}
\lim_{k\rightarrow \infty }\mathcal{R}_{k}(r_{+})=1.
\end{equation}

Equations~\eqref{regular-Psi2} and \eqref{regular-Phi11} describe only the
regular part of the curvature. At $\theta =0$ and $\theta =\pi $, the branch
defects contribute additional delta-function curvature through Eq.~%
\eqref{angular-distributional-curvature}. These distributional terms are not
represented by the ordinary pointwise Newman-Penrose scalars written above.
Consequently, the complete curvature of the branched spacetime consists of
both the regular bulk components and the distributional contributions
supported on the two radial defects. The formal limit $r\rightarrow 0$ is
not relevant to the black-hole exterior considered here, whose radial domain
is $r\geq r_{+}$. Statements about dominance of the $r^{-4}$ term at the
central singularity should therefore not be interpreted as predictions of
the exterior construction.

\subsubsection{Magnitude of the matched bulk energy}

For the minimal normalization $\gamma =1$, the matching condition gives 
\begin{equation}
E_{\mathrm{bulk}}(k)=\frac{4\pi (k-1)\ell _{h}}{\kappa }.
\label{matched-bulk-energy}
\end{equation}%
Restoring SI units through 
\begin{equation}
\kappa =\frac{8\pi G}{c^{4}},
\end{equation}%
we obtain 
\begin{equation}
E_{\mathrm{bulk}}^{(\mathrm{SI})}(k)=\frac{(k-1)c^{4}\ell _{h}}{2G}.
\label{bulk-energy-SI}
\end{equation}%
Using the self-consistent cutoff 
\begin{equation}
\ell _{h}=\frac{2k}{2k-1}\frac{GM}{c^{2}},
\end{equation}%
where $M$ denotes the mass scale associated with the metric parameter $%
m=GM/c^{2}$, Eq.~\eqref{bulk-energy-SI} becomes 
\begin{equation}
E_{\mathrm{bulk}}^{(\mathrm{SI})}(k)=\frac{k(k-1)}{2k-1}\,Mc^{2}.
\label{bulk-energy-mass-ratio}
\end{equation}%
Consequently, 
\begin{equation}
\frac{E_{\mathrm{bulk}}^{(\mathrm{SI})}(k)}{Mc^{2}}=\frac{k(k-1)}{2k-1}.
\end{equation}%
For the first few sectors, 
\begin{equation}
\frac{E_{\mathrm{bulk}}(2)}{Mc^{2}}=\frac{2}{3},\qquad \frac{E_{\mathrm{bulk}%
}(3)}{Mc^{2}}=\frac{6}{5},\qquad \frac{E_{\mathrm{bulk}}(4)}{Mc^{2}}=\frac{12%
}{7}.
\end{equation}%
For large $k$, 
\begin{equation}
E_{\mathrm{bulk}}^{(\mathrm{SI})}(k)\simeq \frac{k}{2}Mc^{2}.
\end{equation}%
The matched bulk energy is therefore not a small quantum correction. It is
comparable to the mass-energy scale of the geometry already for the lowest
nontrivial values of $k$ and grows approximately linearly for large $k$. The
construction should consequently be regarded as a nonperturbative family of
topologically distinct classical geometries, rather than as a spectrum of
small excitations about Schwarzschild spacetime. Moreover, because the
complete branched geometry is not ordinarily asymptotically flat and its
radial defects have divergent absolute energy, $M$ cannot presently be
identified with a globally defined ADM mass of the full spacetime. Equation~%
\eqref{bulk-energy-mass-ratio} should instead be understood as a comparison
with the local mass scale defined by the radial metric parameter $m$.

\section{Weak-field radial acceleration}

\label{sec:weak-field}

Although the construction was developed for the exterior of a branched black
hole, its local weak-field limit can be examined formally away from the
north and south branch defects. This analysis applies to a member of the
present solution family and should not be interpreted as a prediction for an
ordinary material body that has not been shown to possess the same branched
topology. In the weak-field region $r\gg m$, the temporal metric component
may be written as 
\begin{equation}
g_{tt}=-g(r,k)\simeq -\left( 1+\frac{2\Phi _{k}(r)}{c^{2}}\right) .
\end{equation}%
Restoring 
\begin{equation}
m=\frac{GM}{c^{2}},
\end{equation}%
the effective Newtonian potential is 
\begin{equation}
\Phi _{k}(r)=-\frac{GM}{r}+\frac{c^{2}Q_{k}^{2}}{2r^{2}}.
\label{weak-field-potential}
\end{equation}%
The corresponding radial acceleration is 
\begin{equation}
a_{r}=-\frac{d\Phi _{k}}{dr}=-\frac{GM}{r^{2}}+\frac{c^{2}Q_{k}^{2}}{r^{3}}.
\label{weak-field-acceleration}
\end{equation}%
The $Q_{k}$-dependent contribution is directed outward and falls off as $%
r^{-3}$. Relative to the magnitude of the ordinary Newtonian acceleration, 
\begin{equation}
a_{N}=\frac{GM}{r^{2}},
\end{equation}%
the fractional correction is 
\begin{equation}
\epsilon _{k}(r)\equiv \frac{\Delta a_{k}}{a_{N}}=\frac{Q_{k}^{2}}{mr}.
\label{fractional-acceleration}
\end{equation}%
For the minimal normalization $\gamma =1$, 
\begin{equation}
\frac{Q_{k}^{2}}{m^{2}}=\frac{4k(k-1)}{(2k-1)^{2}},
\end{equation}%
and hence 
\begin{equation}
\epsilon _{k}(r)=\frac{4k(k-1)}{(2k-1)^{2}}\frac{m}{r}.
\label{discrete-weak-field-correction}
\end{equation}%
For every finite $k$, 
\begin{equation}
0\leq \frac{4k(k-1)}{(2k-1)^{2}}<1,
\end{equation}%
so that 
\begin{equation}
0\leq \epsilon _{k}(r)<\frac{m}{r}\ll 1
\end{equation}%
throughout the weak-field region. The correction is therefore controlled not
only by the topological degree but also by the usual compactness parameter $%
m/r$. Equation~\eqref{discrete-weak-field-correction} gives the local radial
effect of the smooth bulk term. It does not include the global influence of
the two conical branch defects. Moreover, applying this formula to the
Earth, a star, or another ordinary body would require an independent
argument showing that the body is described by the same branched angular
sector and supporting energy-momentum tensor. No such assumption is made
here. A meaningful phenomenological test would instead require a complete
analysis of orbital motion, light propagation, or tidal observables in the
full branched geometry.

\section{Horizonless zero-mass geometry}

\label{sec:zero-mass}

We next consider the formal zero-mass limit $m=0$. The radial metric
function becomes 
\begin{equation}
g_{k}(r)=1+\frac{Q_{k}^{2}}{r^{2}},
\end{equation}%
and the line element is 
\begin{equation}
ds^{2}=-\left( 1+\frac{Q_{k}^{2}}{r^{2}}\right) dt^{2}+\frac{dr^{2}}{1+%
\dfrac{Q_{k}^{2}}{r^{2}}}+r^{2}f_{k}^{2}(\theta )d\Omega ^{2}.
\label{zero-mass-metric}
\end{equation}%
Since 
\begin{equation}
g_{k}(r)>0\qquad (r>0),
\end{equation}%
this geometry has no Killing horizon. It is therefore a horizonless rather
than a black-hole spacetime. Away from the north and south branch defects,
the smooth bulk energy-momentum tensor is 
\begin{equation}
\left( T^{\mu }{}_{\nu }\right) _{\mathrm{bulk}}
=\frac{Q_{k}^{2}}{\kappa r^{4}}\operatorname{diag}(-1,-1,1,1).
\label{zero-mass-bulk-tensor}
\end{equation}%
Consequently, 
\begin{equation}
\rho _{\mathrm{bulk}}=\frac{Q_{k}^{2}}{\kappa r^{4}}>0,\qquad p_{r,\mathrm{%
bulk}}=-\rho _{\mathrm{bulk}},\qquad p_{\theta ,\mathrm{bulk}}=p_{\varphi ,%
\mathrm{bulk}}=\rho _{\mathrm{bulk}}.
\end{equation}%
The smooth bulk source satisfies the null and weak energy conditions, with 
\begin{equation}
\rho _{\mathrm{bulk}}+p_{r,\mathrm{bulk}}=0,\qquad \rho _{\mathrm{bulk}%
}+p_{\theta ,\mathrm{bulk}}=\rho _{\mathrm{bulk}}+p_{\varphi ,\mathrm{bulk}%
}=2\rho _{\mathrm{bulk}}>0.
\end{equation}%
This statement applies only to the regular bulk source. The distributional
branch defects retain the negative signed tension and energy density
discussed in Sec.~\ref{sec:branch-defects}. The regular Newman-Penrose
curvature scalars are obtained from Eqs.~\eqref{regular-Psi2} and %
\eqref{regular-Phi11} by setting $m=0$: 
\begin{equation}
\Psi _{2}^{(\mathrm{reg})}(r,k)=\frac{Q_{k}^{2}}{r^{4}},\qquad \Phi _{11}^{(%
\mathrm{reg})}(r,k)=\frac{Q_{k}^{2}}{2r^{4}}.  \label{zero-mass-NP-scalars}
\end{equation}%
The regular part of the Kretschmann scalar is 
\begin{equation}
\mathcal{K}_{\mathrm{reg}}\equiv R_{\mu \nu \rho \sigma }R^{\mu \nu \rho
\sigma }=\frac{56Q_{k}^{4}}{r^{8}}.  \label{zero-mass-kretschmann}
\end{equation}%
Thus, $r=0$ is a naked curvature singularity. Introducing a minimum radius
does not remove or regularize this singularity; it merely excises the
singular region from the domain. In the absence of a horizon, the cutoff
scale cannot be fixed by the condition $L_{0}=r_{+}$. We therefore introduce
an independent inner boundary 
\begin{equation}
r\geq L_{0}>0.  \label{zero-mass-cutoff}
\end{equation}%
The relative Killing energy of the smooth bulk source outside this boundary
is 
\begin{equation}
E_{\mathrm{bulk}}^{(0)}(k)=\frac{4\pi kQ_{k}^{2}}{\kappa L_{0}}.
\end{equation}%
If the same phenomenological matching prescription used in the black-hole
sector is imposed, one obtains 
\begin{equation}
\frac{4\pi kQ_{k}^{2}}{\kappa L_{0}}=\frac{4\pi \gamma (k-1)L_{0}}{\kappa },
\end{equation}%
and hence 
\begin{equation}
Q_{k}^{2}=\gamma \left( 1-\frac{1}{k}\right) L_{0}^{2}.
\label{zero-mass-matching}
\end{equation}%
Unlike the black-hole case, however, the geometry itself does not determine $%
L_{0}$. Choosing 
\begin{equation}
L_{0}=\ell _{p}
\end{equation}%
is therefore an additional microscopic assumption rather than a consequence
of the field equations or harmonic-map topology. For the minimal
normalization $\gamma =1$ and the optional choice $L_{0}=\ell _{p}$, the
regular Weyl curvature takes the discrete sequence 
\begin{equation}
\Psi _{2}^{(\mathrm{reg})}(r,k)=\frac{\ell _{p}^{2}}{r^{4}}\left( 1-\frac{1}{%
k}\right) ,\qquad k=1,2,3,\ldots .
\end{equation}%
For $k=2,3,\ldots $, this may be displayed as 
\begin{equation}
\Psi _{2}^{(\mathrm{reg})}=\frac{\ell _{p}^{2}}{r^{4}}\left( \frac{1}{2},%
\frac{2}{3},\frac{3}{4},\ldots ,\frac{n}{n+1},\ldots \right) ,
\end{equation}%
with 
\begin{equation}
\Phi _{11}^{(\mathrm{reg})}=\frac{1}{2}\Psi _{2}^{(\mathrm{reg})}.
\end{equation}%
These are discrete values of the regular curvature at fixed $r$; they should
not be interpreted as quantum-mechanical orbital energy levels. The local
weak-field potential of the smooth bulk term is 
\begin{equation}
\Phi _{k}(r)=\frac{c^{2}Q_{k}^{2}}{2r^{2}},
\end{equation}%
which produces the outward radial acceleration 
\begin{equation}
a_{r}=\frac{c^{2}Q_{k}^{2}}{r^{3}}.
\end{equation}%
This repulsive local acceleration does not make the source equivalent to
dark energy. The bulk tensor is anisotropic, traceless, and decays as $%
r^{-4} $, whereas a cosmological constant has 
\begin{equation}
T^{\mu }{}_{\nu }=-\rho _{\Lambda }\delta ^{\mu }{}_{\nu }
\end{equation}%
with constant $\rho _{\Lambda }$. No dark-energy interpretation is therefore
assigned to the zero-mass solution. Finally, the excised geometry with $%
r\geq L_{0}$ is not a complete regular spacetime unless an interior geometry
and appropriate junction conditions are supplied at $r=L_{0}$. We leave the
construction of such an interior completion for future study.

\section{Simpson-Visser geometry and a limitation of the matching
prescription}

\label{sec:simpson-visser}

We next apply the branched angular construction to the Simpson--Visser
geometry. Let 
\begin{equation}
R(r)=\sqrt{r^{2}+\ell _{0}^{2}},\qquad \ell _{0}>0,  \label{SV-areal-radius}
\end{equation}%
and consider 
\begin{equation}
ds^{2}=-F_{k}(r)dt^{2}+\frac{dr^{2}}{F_{k}(r)}+R^{2}(r)f_{k}^{2}(\theta
)d\Omega ^{2},  \label{branched-SV-metric}
\end{equation}%
where 
\begin{equation}
F_{k}(r)=1-\frac{2m}{R(r)}+\frac{Q_{k}^{2}}{R^{2}(r)}.
\label{SV-metric-function}
\end{equation}%
For $k=1$ and $Q_{1}=0$, this reduces to the original Simpson-Visser
geometry \cite{simpson2019}. Depending on the values of $m$, $\ell _{0}$,
and $Q_{k}$, the metric may represent a traversable wormhole, a black
bounce, or a limiting one-way geometry. The throat-to-infinity
Killing-energy calculation below applies directly to the static
traversable-wormhole regime, in which 
\begin{equation}
F_{k}(r)>0\qquad (-\infty <r<\infty ).
\end{equation}%
The angular factor $f_{k}^{2}(\theta )$ produces the same north and south
branch defects found in the Schwarzschild construction. Their distributional
source must again be separated from the regular energy-momentum tensor. In
addition, the $Q_{k}$-dependent part of the regular source cannot be
obtained by simply replacing $r$ with $R$ in the RN
tensor, because 
\begin{equation}
R^{\prime }(r)=\frac{r}{R},\qquad R^{\prime \prime }(r)=\frac{\ell _{0}^{2}}{%
R^{3}}
\end{equation}%
contribute to the Einstein tensor. To isolate the effect of $Q_{k}$, we
subtract the regular Einstein tensor of the $Q_{k}=0$ Simpson-Visser
geometry while keeping $m$, $\ell _{0}$, and $k$ fixed: 
\begin{equation}
\Delta G^{\mu }{}_{\nu }\equiv G^{\mu }{}_{\nu }(Q_{k})-G^{\mu }{}_{\nu }(0).
\end{equation}%
A direct calculation gives 
\begin{equation}
\Delta G^{t}{}_{t}=\frac{Q_{k}^{2}(2\ell _{0}^{2}-r^{2})}{(r^{2}+\ell
_{0}^{2})^{3}},  \label{SV-delta-Gtt}
\end{equation}%
\begin{equation}
\Delta G^{r}{}_{r}=-\frac{Q_{k}^{2}r^{2}}{(r^{2}+\ell _{0}^{2})^{3}},
\label{SV-delta-Grr}
\end{equation}%
and 
\begin{equation}
\Delta G^{\theta }{}_{\theta }=\Delta G^{\varphi }{}_{\varphi }=\frac{%
Q_{k}^{2}r^{2}}{(r^{2}+\ell _{0}^{2})^{3}}.  \label{SV-delta-Gangular}
\end{equation}%
The corresponding relative bulk energy density and pressures are therefore 
\begin{equation}
\Delta \rho _{k}=-\Delta T^{t}{}_{t}=\frac{Q_{k}^{2}(r^{2}-2\ell _{0}^{2})}{%
\kappa (r^{2}+\ell _{0}^{2})^{3}},  \label{SV-relative-density}
\end{equation}%
\begin{equation}
\Delta p_{r,k}=-\frac{Q_{k}^{2}r^{2}}{\kappa (r^{2}+\ell _{0}^{2})^{3}},
\end{equation}%
and 
\begin{equation}
\Delta p_{\theta ,k}=\Delta p_{\varphi ,k}=\frac{Q_{k}^{2}r^{2}}{\kappa
(r^{2}+\ell _{0}^{2})^{3}}.
\end{equation}%
Thus, the $Q_{k}$-dependent relative energy density is negative near the
throat, 
\begin{equation}
\Delta \rho _{k}(0)=-\frac{2Q_{k}^{2}}{\kappa \ell _{0}^{4}},
\end{equation}%
changes sign at 
\begin{equation}
|r|=\sqrt{2}\,\ell _{0},
\end{equation}%
and becomes positive in the asymptotic regions. For one asymptotic branch, $%
0\leq r<\infty $, the relative Killing energy is 
\begin{equation}
\Delta E_{\mathrm{bulk}}^{(+)}(k)=\int_{\Sigma _{t}^{(+)}}\Delta T_{\mu \nu
}n^{\mu }\xi ^{\nu }\sqrt{h}\,d^{3}x=\frac{4\pi kQ_{k}^{2}}{\kappa }%
\int_{0}^{\infty }\frac{r^{2}-2\ell _{0}^{2}}{(r^{2}+\ell _{0}^{2})^{2}}\,dr.
\label{SV-energy-integral}
\end{equation}%
The radial integral is 
\begin{equation}
\int_{0}^{\infty }\frac{r^{2}-2\ell _{0}^{2}}{(r^{2}+\ell _{0}^{2})^{2}}%
\,dr=-\frac{\pi }{4\ell _{0}},
\end{equation}%
and hence 
\begin{equation}
\Delta E_{\mathrm{bulk}}^{(+)}(k)=-\frac{\pi ^{2}kQ_{k}^{2}}{\kappa \ell _{0}%
}.  \label{SV-one-branch-energy}
\end{equation}%
Including both asymptotic branches gives 
\begin{equation}
\Delta E_{\mathrm{bulk}}^{(\mathrm{two})}(k)=-\frac{2\pi ^{2}kQ_{k}^{2}}{%
\kappa \ell _{0}}.  \label{SV-two-branch-energy}
\end{equation}%
This result reveals a limitation of the energy-matching prescription. For $%
\gamma >0$, the scaled excess harmonic-map energy on one branch, 
\begin{equation}
\Delta E_{k}^{(\mathrm{map})}=\frac{4\pi \gamma (k-1)\ell _{0}}{\kappa },
\end{equation}%
is positive for $k>1$, whereas Eq.~\eqref{SV-one-branch-energy} is negative
for real nonzero $Q_{k}$. Consequently, the condition 
\begin{equation}
\Delta E_{\mathrm{bulk}}^{(+)}(k)=\Delta E_{k}^{(\mathrm{map})}
\end{equation}%
has no solution with 
\begin{equation}
Q_{k}^{2}>0,\qquad \gamma >0.
\end{equation}%
The relation 
\begin{equation}
Q_{k}^{2}=\left( 1-\frac{1}{k}\right) \ell _{0}^{2}
\end{equation}%
obtained by treating $R$ as an independent integration coordinate is
therefore not valid for the Simpson-Visser metric.

The harmonic-map construction still generates a well-defined branched
Simpson-Visser geometry through the angular factor $f_{k}^{2}(\theta )$.
However, the particular positive bulk-energy matching used in the
Schwarzschild sector does not extend to the radial ansatz in Eq.~%
\eqref{SV-metric-function}. Achieving such a matching would require a
different radial deformation, a different sign or coupling prescription, or
a dynamical action that determines the interaction between the harmonic-map
and wormhole sectors. We do not impose such an additional modification here.

\section{de Sitter geometry and a common-region vacuum-energy comparison}

\label{sec:de-sitter}

We finally apply the branched angular construction to the static de Sitter
geometry. Consider 
\begin{equation}
ds^{2}=-F_{k}(r)dt^{2}+\frac{dr^{2}}{F_{k}(r)}+r^{2}f_{k}^{2}(\theta
)d\Omega ^{2},  \label{branched-de-Sitter-metric}
\end{equation}%
where 
\begin{equation}
F_{k}(r)=1-\frac{r^{2}}{ka^{2}}.  \label{de-Sitter-function}
\end{equation}%
For the reference sector $k=1$, define 
\begin{equation}
\Lambda =\frac{3}{a^{2}}.
\end{equation}%
Away from the north and south branch defects, the degree-$k$ geometry is
locally de Sitter with the effective cosmological constant 
\begin{equation}
\Lambda _{k}=\frac{\Lambda }{k}=\frac{3}{ka^{2}}.  \label{effective-Lambda-k}
\end{equation}%
Accordingly, its regular Einstein tensor and vacuum energy-momentum tensor
are 
\begin{equation}
\left( G^{\mu }{}_{\nu }\right) _{\mathrm{reg}}=-\Lambda _{k}\delta ^{\mu
}{}_{\nu },
\end{equation}%
and 
\begin{equation}
\left( T^{\mu }{}_{\nu }\right) _{\Lambda ,k}=-\frac{\Lambda _{k}}{\kappa }%
\delta ^{\mu }{}_{\nu },
\end{equation}%
respectively. The regular vacuum-energy density and scalar curvature are
therefore 
\begin{equation}
\rho _{\Lambda ,k}=\frac{\Lambda _{k}}{\kappa },\qquad R_{\mathrm{reg}%
,k}=4\Lambda _{k}=\frac{12}{ka^{2}}.  \label{de-Sitter-density-curvature}
\end{equation}%
These expressions apply away from the branch defects, whose distributional
curvature and negative signed tension remain present as in Sec.~\ref%
{sec:branch-defects}. The cosmological Killing horizon of the $k$-sector is
determined by $F_{k}(r_{c,k})=0$ and is located at 
\begin{equation}
r_{c,k}=\sqrt{k}\,a.  \label{k-cosmological-horizon}
\end{equation}%
Thus, different values of $k$ correspond not merely to different angular
coverings but also to different effective cosmological constants and
different static-patch radii. They should consequently be regarded as
solutions belonging to different effective-vacuum sectors, rather than
solutions of a single Einstein equation with one fixed value of $\Lambda $.
To compare the regular vacuum energies, first consider the common coordinate
region 
\begin{equation}
\mathcal{B}_{a}:\qquad 0\leq r\leq a.
\end{equation}%
This region fills the $k=1$ static patch and lies inside the cosmological
horizon for every $k>1$. The Killing energy of the degree-$k$ vacuum in this
region is 
\begin{equation}
E_{\Lambda ,k}(\mathcal{B}_{a})=\int_{\mathcal{B}_{a}}\left( T_{\mu \nu
}\right) _{\Lambda ,k}n^{\mu }\xi ^{\nu }\sqrt{h}\,d^{3}x=\frac{\Lambda _{k}%
}{\kappa }\int_{0}^{a}r^{2}\,dr\int_{S^{2}}f_{k}^{2}(\theta )\,d\Omega =%
\frac{4\pi a}{\kappa }.  \label{actual-de-Sitter-energy}
\end{equation}%
For $k=1$, the corresponding result is 
\begin{equation}
E_{\Lambda ,1}(\mathcal{B}_{a})=\frac{4\pi a}{\kappa }.
\end{equation}%
Hence, the direct difference between the actual regular vacuum energies in
the common region is 
\begin{equation}
E_{\Lambda ,k}(\mathcal{B}_{a})-E_{\Lambda ,1}(\mathcal{B}_{a})=0.
\label{direct-de-Sitter-difference}
\end{equation}%
The decrease of the vacuum-energy density is exactly compensated by the
increase of the branched angular volume. A nonzero quantity arises if both
vacuum densities are instead compared on the same $k$-branched spatial
volume. Define the reference energy 
\begin{equation}
E_{\Lambda }^{\mathrm{ref},k}(\mathcal{B}_{a})=\frac{\Lambda }{\kappa }%
\int_{0}^{a}r^{2}\,dr\int_{S^{2}}f_{k}^{2}(\theta )\,d\Omega =\frac{4\pi ka}{%
\kappa }.  \label{de-Sitter-reference-energy}
\end{equation}%
The reduction of this reference energy when $\Lambda \rightarrow \Lambda
_{k}=\Lambda /k$ is 
\begin{align}
\Delta E_{\Lambda }^{\mathrm{ref}}(k)& \equiv E_{\Lambda }^{\mathrm{ref},k}(%
\mathcal{B}_{a})-E_{\Lambda ,k}(\mathcal{B}_{a})  \notag \\
& =\frac{4\pi (k-1)a}{\kappa }.  \label{de-Sitter-reference-reduction}
\end{align}%
For the minimal normalization $\gamma =1$, the scaled excess harmonic-map
energy is 
\begin{equation}
\Delta E_{k}^{(\mathrm{map})}=\frac{a}{\kappa }\left[ E_{k}^{(\mathrm{map}%
)}-E_{1}^{(\mathrm{map})}\right] =\frac{4\pi (k-1)a}{\kappa }.
\label{de-Sitter-map-energy}
\end{equation}%
Consequently, 
\begin{equation}
\Delta E_{\Lambda }^{\mathrm{ref}}(k)=\Delta E_{k}^{(\mathrm{map})}.
\label{de-Sitter-reference-matching}
\end{equation}%
This equality is a common-volume reference matching: it compares the
original and reduced vacuum densities on the same degree-$k$ branched
geometry. It should not be interpreted as the direct energy difference
between the physical $k$ and $k=1$ sectors, which vanishes in the region $%
\mathcal{B}_{a}$ according to Eq.~\eqref{direct-de-Sitter-difference}.

For completeness, the regular vacuum energy of each sector in its own
complete static patch is 
\begin{align}
E_{\Lambda ,k}^{(\mathrm{patch})}& =\frac{\Lambda _{k}}{\kappa }%
\int_{0}^{r_{c,k}}r^{2}\,dr\int_{S^{2}}f_{k}^{2}(\theta )\,d\Omega  \notag \\
& =\frac{4\pi k^{3/2}a}{\kappa }.  \label{full-de-Sitter-patch-energy}
\end{align}%
Thus, 
\begin{equation}
E_{\Lambda ,k}^{(\mathrm{patch})}-E_{\Lambda ,1}^{(\mathrm{patch})}=\frac{%
4\pi a}{\kappa }\left( k^{3/2}-1\right) ,
\end{equation}%
which does not equal the excess harmonic-map energy. The equality in Eq.~%
\eqref{de-Sitter-reference-matching} therefore depends essentially on the
common-region reference prescription. The de Sitter example consequently
provides a consistent algebraic normalization of the map-energy difference
through a reduction of the vacuum density on a fixed branched volume. It
does not show that the cosmological constant dynamically supplies the map
energy. Establishing such a mechanism would require a theory in which $%
\Lambda $ is dynamical and coupled to the harmonic-map sector.

\section{Conclusion}

We have investigated a topological construction of general-relativistic
geometries based on harmonic self-maps of the two-sphere. The integer degree 
$k$ labels disconnected topological sectors, but for $k>1$ the pullback
angular metric is not globally smooth. It defines a branched covering with
conical excesses at the north and south poles. These branch points extend
over the $(t,r)$ sector as codimension-two radial defects with negative
signed tension and distributional curvature. The resulting geometries are
locally equivalent to their unbranched counterparts away from the defects
but have different global properties. For the branched Schwarzschild
geometry, the RN-like term $Q_{k}^{2}/r^{2}$ generates a
smooth bulk energy-momentum tensor in addition to the distributional defect
source. Its finite Killing energy is defined relative to the $Q_{k}=0$
branched geometry at fixed $k$. We imposed a phenomenological matching
between this relative bulk energy and the scaled excess harmonic-map energy.
For the minimal normalization $\gamma =1$, the additional condition $\ell
_{h}=r_{+}$ determines both the matching scale and $Q_{k}$. The resulting
sequence begins with Schwarzschild at $k=1$ and approaches an extremal
RN-like limit as $k\rightarrow \infty $. The horizon
area and Hawking temperature are geometrically well defined. The entropy
retains the area form only under the stated assumption that the defects are
fixed external sources without curvature-dependent interactions. A complete
first law would require well-defined global Hamiltonian charges and a
possible work term associated with the defect tension. The regular
Newman-Penrose scalars and the local weak-field acceleration acquire
explicit $k$-dependence, although the weak-field correction remains bounded
by the compactness parameter $m/r$. In the zero-mass limit, the construction
gives a horizonless geometry with a positive-energy smooth bulk source,
negative-energy branch defects, and a naked central curvature singularity.
Introducing an inner boundary excises rather than resolves this singularity
and requires an additional interior completion. The Simpson-Visser example
demonstrates that the positive Schwarzschild matching prescription is not
universal. For the radial ansatz considered here, the $Q_{k}$-dependent
relative Killing energy is negative and cannot be matched to the positive
excess map energy with real $Q_{k}$ and positive coupling. In the de Sitter
construction, the decrease of the regular vacuum density is compensated by
the increase of the branched angular volume in a direct common-region
comparison. The excess map energy is recovered only through a reference
subtraction performed on the same $k$-branched volume, rather than through a
dynamical transfer of vacuum energy. The present construction is classical
and should not be interpreted as a complete theory of quantum gravity. Its
robust result is the existence of integer-labeled branched geometrical
sectors and their associated distributional defects. The bulk-energy
matching is an additional prescription whose validity depends on the radial
geometry. Deriving such a relation dynamically would require a
four-dimensional action coupling the harmonic-map degrees of freedom, the
defect sources, and the spacetime metric.

%%%%%%%%%%%%%%%%%%%%%%%%%%%%%%%%%%%%%%%%%%%%%%%%%%%%%%%%%%%%%%%%%%%%%%%%%%%

\end{document}